\documentclass[notitlepage,prd,twocolumn,showpacs,preprintnumbers,nofootinbib,superscriptaddress]{revtex4-1}
\usepackage{amssymb,amsmath,placeins,epsfig,array}
\usepackage[usenames,dvipsnames]{color}
\usepackage[normalem]{ulem}

\usepackage{graphicx,slashed,hyperref}
\hypersetup{urlcolor=blue, citecolor=blue, colorlinks=true, breaklinks, colorlinks}
\pdfoutput=1
\usepackage{xcolor}

\def\gsim{\mathrel{\raise.3ex\hbox{$>$\kern-.75em\lower1ex\hbox{$\sim$}}}}

\newcommand{\ie}{{\it i.e.~}}  \newcommand{\eg}{{\it e.g.~}}

\newcommand{\beq}{\begin{equation}} \newcommand{\eeq}{\end{equation}}
\newcommand{\bea}{\begin{eqnarray}} \newcommand{\eea}{\end{eqnarray}}

\newcommand{\Eq}[1]{Eq.~(\ref{#1})}  
\newcommand{\Sec}[1]{Sec.~\ref{#1}}  
\newcommand{\Fig}[1]{Fig.~\ref{#1}}

\begin{document}

\title{Imprints of Axion Superradiance in the CMB}
\author{Diego Blas}\email{e-mail: diego.blas@kcl.ac.uk}
\affiliation{Theoretical Particle Physics and Cosmology Group, Department of Physics,\\
  ~~ King's College London, Strand, London WC2R 2LS, UK}
\author{Samuel J.~Witte}\email{e-mail: witte.sam@gmail.com}
\affiliation{Instituto de F\'{\i}sica Corpuscular (IFIC), CSIC-Universitat de Val\`encia, Spain}
\preprint{KCL-2020-56}

\begin{abstract}
Light axions ($m_a \lesssim 10^{-10}$ eV) can form dense clouds around rapidly rotating astrophysical black holes via a mechanism known as rotational superradiance. The coupling between axions and photons induces a parametric resonance, arising from the stimulated decay of the axion cloud, which can rapidly convert regions of large axion number densities into an enormous flux of low-energy photons. In this work we consider the phenomenological implications of a superradiant axion cloud undergoing resonant decay. We show that the low energy photons produced from such events will be absorbed over cosmologically short distances, potentially inducing massive shockwaves that heat and ionize the IGM over Mpc scales. These shockwaves may leave observable imprints in the form of anisotropic spectral distortions or inhomogeneous features in the optical depth.
\end{abstract}
\maketitle

\section{Introduction}

The evidence for the existence of dark matter is abundant in both cosmology and astrophysics, however its underlying nature has evaded detection efforts for decades. The axion, a pseudo goldstone boson arising from the breaking of a global $U_{PQ}(1)$ symmetry,  has recently emerged as one of the most favored dark matter candidates. Originally introduced as a solution to the strong-CP problem~\cite{Peccei:1977hh,Peccei:1977ur,Weinberg:1977ma,Wilczek:1977pj}, it was soon realized that light axions $m_a \lesssim 10^{-3}$ eV were both cosmologically stable and could be abundantly produced in the early Universe via the misalignment mechanism~\cite{Preskill:1982cy,Dine:1982ah,Abbott:1982af,Davis:1986xc}. Ultralight bosonic fields have also been predicted to naturally appear from moduli compactification in string theory~\cite{Arvanitaki:2009fg,Marsh:2015xka}; while these particles do not solve the strong-CP problem, they behave in a comparable manner to the QCD axion, and thus are frequently referred to as axion-like particles. In what follows, we will use the word `axion' interchangeably to refer to both.

Light axions with masses $m_a \lesssim 10^{-10}$ eV are known to trigger instabilities around rotating astrophysical black holes, even if their initial abundance is negligibly small. This processes, known as black hole superradiance, produces an exponentially growing boson cloud at the expense of the black hole's rotational energy~\cite{zel1971generation,Press:1972zz,Teukolsky:1974yv,Damour:1976kh,detweiler1980klein,Cardoso:2004nk} (see~\cite{Brito:2015oca} for a recent review on rotational superradiance and its application to black hole physics).  In the limit that the axion's interactions can be neglected (that is, the limit that the axion's equation of motion is governed by the Klein-Gordon equation), the axion cloud will grow until either the superradiant condition, given by $\omega < m \Omega$ -- with $\omega$ the energy of the mode, $m$ the azimuthal quantum number and $\Omega$ the rotational frequency-- is violated or a significant fraction of the black hole's rotational energy has been extracted\footnote{Current work suggests the appearance and growth of the instability is robust against gravitational emission, back-reaction on the metric, and the presence of accretion disks~\cite{East:2013mfa,Brito:2014wla}.}; this cloud will then slowly decay via the emission of gravitational waves, with amplitudes that may be detectable in future experiments~\cite{Arvanitaki:2010sy,Arvanitaki:2014wva,Arvanitaki:2016qwi,Baryakhtar:2017ngi,Brito:2017wnc,Brito:2017zvb}. Observations of highly rotating black holes have consequently been used to probe and constrain light non-interacting bosons with masses in the range $10^{-21} \lesssim m_\phi \lesssim 10^{-10}$ eV~\cite{Arvanitaki:2009fg,Arvanitaki:2010sy,Arvanitaki:2014wva,Brito:2017zvb,Brito:2017wnc,Rosa:2017ury,Zhu:2020tht,Davoudiasl:2019nlo}.

The existence of self-interactions or couplings to other particles may complicate and quench the superradiant growth; for example, in the case of the axion, self-interactions become important when the field value $a$ approaches the axion decay constant $f_a$, and may produce an explosion known as a `bosenova'~\cite{Arvanitaki:2009fg,Arvanitaki:2010sy}. Particle production and scattering processes may dramatically increase the characteristic timescale required to extract the black hole's angular momentum, and thus provide a natural way for evading current superradiant constraints~\cite{Fukuda:2019ewf,Mathur:2020aqv,BlasWitte_photonSR}. More recently, it has been pointed out that the parametric resonance arising from the repeated stimulated decay of a superradiant axion cloud may quench the growth, producing in the process an enormous flux of low energy photons~\cite{Rosa:2017ury,Sen:2018cjt,Boskovic:2018lkj,Ikeda:2018nhb}. While initial studies of this phenomenon assumed adiabatic evolution in a flat-spacetime, recent work has demonstrated that axion fields coupled to electromagnetism in Kerr background robustly exhibit this phenomenon, so long as the axion-photon coupling exceeds a threshold value~\cite{Boskovic:2018lkj,Ikeda:2018nhb}. 

In this work we investigate the phenomenological implications that would arise from the resonant decay of a superradiant axion cloud. We show that the low energy photons produced in the resonant decay have short cosmological mean free paths, and are typically absorbed over $\sim \mathcal{O}({\rm pc})$ scales. This leads to enormous pressure gradients, which can induce a shockwave thats heats and ionizes the surrounding medium as the shock front pushes outward. We show that these shockwaves leave anisotropic features in the spectrum and optical depth of the Cosmic Microwave Background (CMB); individual features for a single black  hole may be detectable with near-future spectral distortion experiments, while large-scale features arising from the cumulative distribution of black holes may be near current detection thresholds, however this latter point requires a more dedicated study.

This manuscript is organized as follows. We review the phenomenon of axion induced black hole superradiance in \Sec{sec:superR}, including possible outcomes (specifically we focus on the possibility of either a bosenova or photon production via resonant decay). In \Sec{sec:SW} we discuss the propagation of the low energy photons in the IGM, and the subsequent formation and properties of the shockwave. Observable features arising from such shockwaves are discussed in \Sec{sec:obs}. We conclude in \Sec{sec:conc}.

\section{Axion Superradiance}\label{sec:superR}

Rotational superradiance is a phenomenon that occurs when a low energy boson with energy $\omega < m \Omega$ scatters off an absorbing surface rotating with frequency $\Omega$ ($m$ here being the azimuthal quantum number). In this scattering process, the amplitude of the reflected boson field is enhanced with respect to the incident amplitude. Should there exist a confining mechanism for the reflected radiation, this process will extract the rotational energy at an exponential rate until the growth is quenched by additional interactions or the superradiant condition is no longer satisfied. 

In the context of black hole superradiance, a non-zero particle mass naturally acts to confine the light boson around the black hole, and exponential growth can occur. The timescale of the instability is directly related to the mass of boson $m_a$ and the black hole $M$, and for the case of scalar particles can be as large as (in the limit ${G} \, m_a M \ll 1$\footnote{{We work here with $\hbar = c = 1$.}}) ~\cite{Zouros:1979iw,Detweiler:1980uk,Arvanitaki:2010sy,Witek:2012tr}
\begin{equation}
	\tau_{sr} \sim 10^{2} \, \left( \frac{M}{10 \, M_\odot}\right) \, \left(\frac{0.2}{{G} \, m_a \, M}\right)^9 \, s \, .
\end{equation}
The size of the generated axion cloud is approximately given by
\begin{equation}
	r_{\rm cloud} \sim \frac{(\ell + n + 1)^2 \, {G} \, M}{({G} \, m_a M)^2} \, ,
\end{equation}
which extends well beyond the horizon where rotational effects can be neglected.

\begin{figure*}
	\includegraphics[width=0.48\textwidth, bb=0 0 625 400]{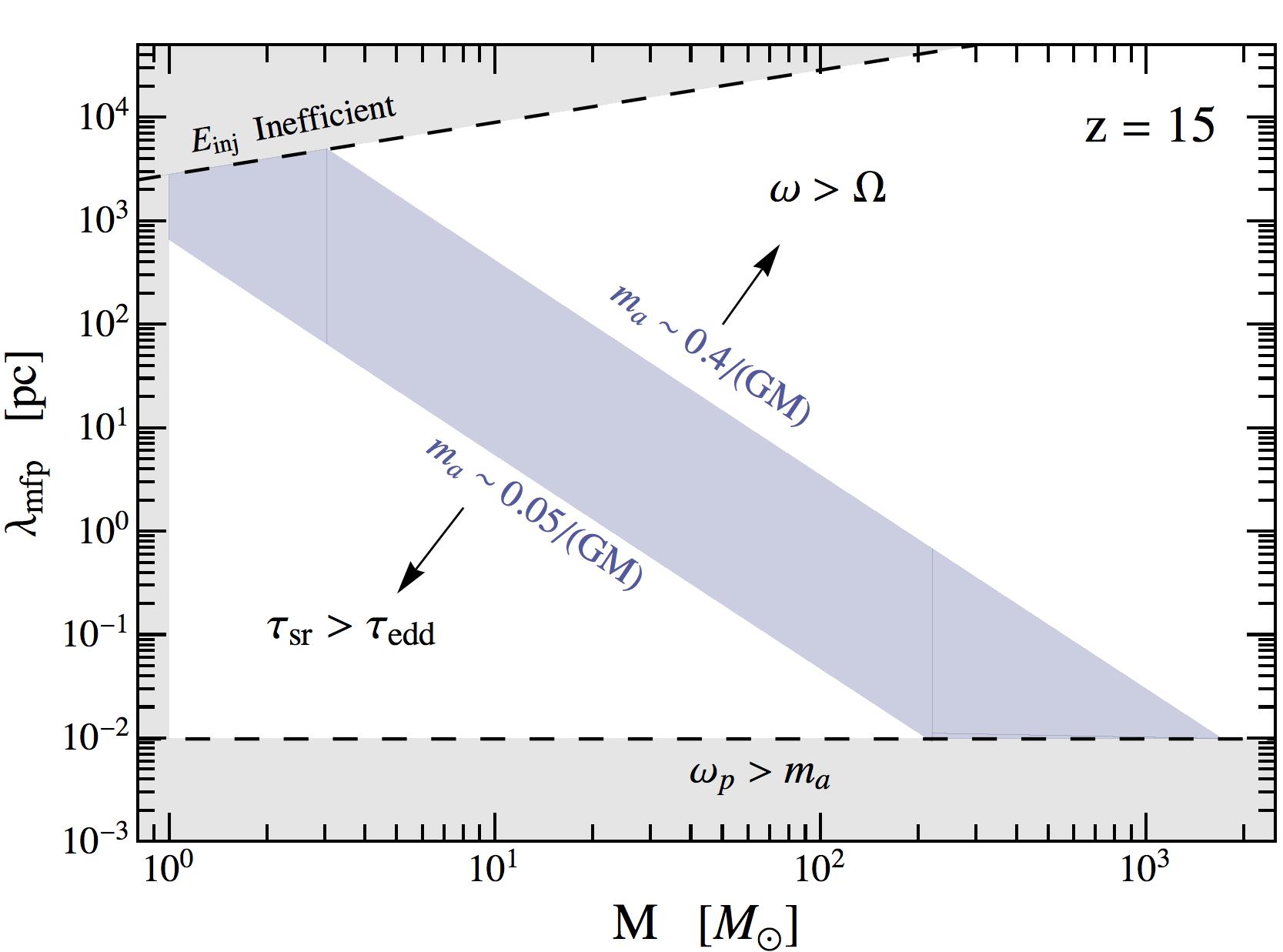}
    \includegraphics[width=0.48\textwidth,  bb=0 0 625 400]{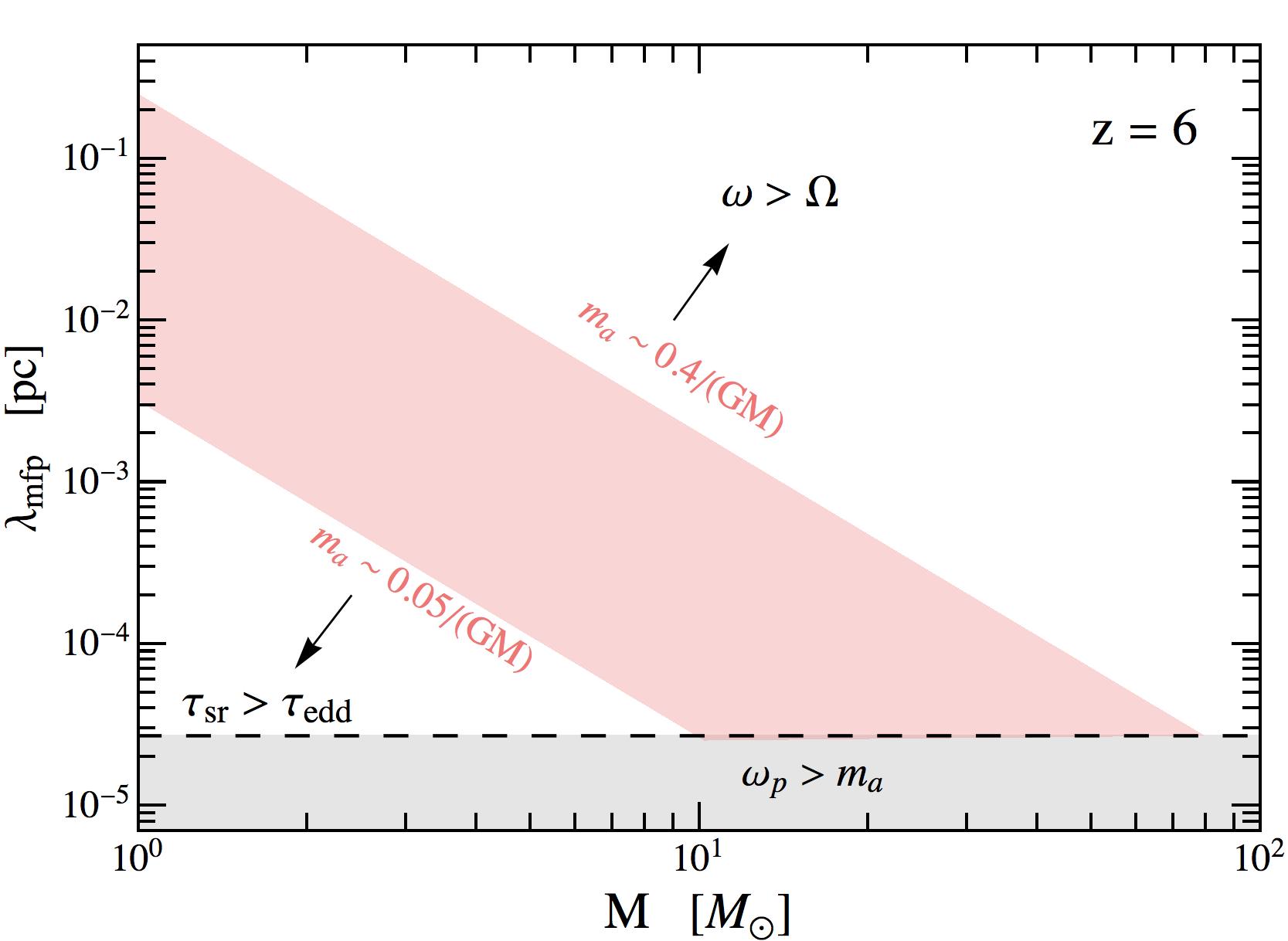}
    \caption{\label{fig:mfp} Mean free path of photons produced from the resonant decay of superradiant axions of mass $m_a$ near a black hole of mass $M$, assumed to occur at $z = 15$ (left) or $z = 6$ (right). Interesting regions of parameter space are identified in color, while regions where energy injection is inefficient or impossible are heuristically identified. }
\end{figure*}

Assuming the axion potential is given by the conventional
\begin{equation}
	V(\phi) = m_a^2 \, f_a^2 \left[ 1 - \cos\left(\frac{a}{f_a}\right)\right] \, ,
\end{equation}
non-linear self-interactions of the axion may become important when the mass contained in the axion cloud $M_a$ is~\cite{Kodama:2011zc,Yoshino:2012kn,Yoshino:2014wwa}
\begin{equation}\label{eq:bosenova}
	M_a \sim 1600\, (G\, M)  \, f_a^2 \, ;
\end{equation}
we emphasize that this relation was computed for ${G} \, m_a M \sim 0.4$ only, and the scaling with this parameter has not yet been rigorously tested. Should the axion field grow to this point, a bosenova may ensue in which axions are partially ejected and partially absorbed by the black hole. Axions, however, generically couple to electromagnetism via
\begin{equation}
	\mathcal{L} \supset - \frac{g_{a \gamma \gamma}}{2} \, a \, F^{\mu\nu} \, \tilde{F}_{\mu\nu} \, ,
\end{equation}
where $F$ and $\tilde{F}$ are the electromagnetic field strength tensor and its dual; this coupling generates a two-body decay to photons, which in the case of non-relativistic axions are approximately monochromatic with energies $E_\gamma \sim m_a / 2$. If the axion field is sufficiently dense and the energy distribution sufficiently narrow, the production of photons at this frequency via axion decays may stimulate the decay of the other axions in the cloud~\cite{Tkachev:1987cd,Tkachev:2014dpa,Arza:2018dcy,Hertzberg:2018zte,Sigl:2019pmj,Alonso-Alvarez:2019ssa,Arza:2020eik,Levkov:2020txo}\footnote{It is worth mentioning that the difference between the resonant decay of the axion field (\ie the focus of this work) and the stimulated decay discussed in ~\cite{Caputo:2018ljp,Caputo:2018vmy}, is that the latter does not assume photons produced from the decays of axions contribute to the photon occupation number responsible for stimulating the decay; rather, the photon occupation number is assumed only to arise from CMB photons, galactic synchrotron and free-free emission, etc. The approach of~\cite{Caputo:2018ljp,Caputo:2018vmy} thus represents the conservative limit of the resonant decay process. }. This processes is typically suppressed in conventional astrophysical and cosmological environments, due to either the finite width of the axion momentum distribution or gravitational de-tuning (see \eg~\cite{Alonso-Alvarez:2019ssa,Arza:2020eik}); superradiant axion fields naturally evade these concerns, however, due to their enormous densities. Should the resonant decay of a superradiant axion cloud occur, an enormous burst of monochromatic low-energy photons will be emitted that will propagate away from the black hole and into the IGM. 

It has recently been shown numerically that an axion field coupled to electromagnetism in a Kerr metric does indeed induce this parametric resonant decay, but only if the axion-photon coupling is sufficiently large. The condition derived on axion-photons couplings for which this occurs is given by~\cite{Boskovic:2018lkj,Ikeda:2018nhb}
\begin{equation}\label{eq:ax_coup}
	g_{a\gamma\gamma } \, \gtrsim  \, 10^{-19} \, \sqrt{\frac{M}{M_a}} \, \frac{1}{({G} \, m_a \, M)^2} \, {\rm GeV^{-1}} \, .
\end{equation}
One can compare the fractional mass in the axion cloud at the time of a bosenova with that at the time of the electromagnetic burst (\ie \Eq{eq:bosenova} with \Eq{eq:ax_coup}); assuming $f_a \sim 1 / g_{a\gamma\gamma}$\footnote{In general there also exists an additional model-dependent pre-factor relating $f_a$ to $g_{a\gamma\gamma}$ which may significantly alter this relation. For the QCD axion, relation changes to $f_a = \alpha /(4\pi) \times (E/N) $ , with $E$ and $N$ given by anomaly coefficients (see \eg~\cite{Irastorza:2018dyq}). } one finds the axion decay occurs prior to the bosenova if 
\begin{equation}
	2\times 10^{3} \,  (G \, m_a \, M)^4 \gtrsim 1 \, .
\end{equation}
This relation is only valid at  $({G} \, m_a M) \sim 0.4$  (note for a $1 \, M_\odot$ black hole, this choice corresponds to an axion mass of $m_a \sim 6 \times 10^{-11}$ eV) as \Eq{eq:bosenova} has only been derived at this value -- here, one finds the condition is, at least naively, safely satisfied. The peak luminosity was estimated numerically, and was found (assuming $m_a \sim 0.2/({G} \, M)$ and a coupling $g_{a\gamma\gamma}$ near the threshold value) to be roughly~\cite{Ikeda:2018nhb}
\begin{equation}
	\frac{dE}{dt} \sim 10^{66} \, \left(\frac{M_a}{M} \right) \, {\rm eV/s} \, .
\end{equation}
This emission is expected to occur in bursts; while the properties of the bursts are not known, we can approximate the  time between bursts as the timescale required to replenish the axion cloud $\tau_{b}\sim 10^2 \tau_{sr}$, and the duration of the bursts $\tau_{\Delta} \sim \mathcal{O}({\rm ms})$~\cite{Rosa:2017ury}. We caution the reader that large uncertainties are associated with both of these numbers. We can also estimate the average luminosity over timescales $t \gg \tau_{ b}$, which is roughly given by
\begin{equation}\label{eq:timeAL}
	\overline{L} \sim \left(\frac{\tau_{\Delta}}{\tau_{b}} \right) L \, .
\end{equation}

The size of the axion cloud when these bursts occur has also not been robustly determined. We can, however, determine an upper limit by requiring  that this emission does not significantly change the total black hole spin -- in this way, this axion candidate can evade current superradiant constraints. We do this by imposing that  the time-integrated luminosity does not exceed a fraction $\beta$ of the black hole's rotational energy -- for the results presented here, we will fix $\beta = 0.5$ in order to optimistically assess the detection thresholds. In general, axion superradiance is expected to occur until the black hole mass grows to a point where either the superradiant condition is no longer valid or the timescale for superradiance is longer than the age of the Universe. The most likely process for changing the black hole mass is accretion.  The minimum timescale over which accretion is expected to change the black hole mass is given by $\tau_{edd}$, \ie the characteristic timescale for Eddington limited accretion, given by~\cite{Brito:2015oca}
\begin{equation}
	\tau_{edd} \simeq \frac{M}{\dot{M}} \sim \frac{\epsilon \sigma_T}{4\pi G m_p} \sim 1.4 \, \epsilon \, \times 10^{15} \, {\rm s}.
\end{equation}
Here, $\epsilon$ is the radiative efficiency of the black hole, taken here to be $\sim 0.3$~\cite{Thorne:1974ve}. The constraint on the the mass of the axion cloud is thus given by
\begin{equation}
	\left( \frac{M_a}{M} \right) \lesssim \,  \beta \, \left( \frac{M}{M_\odot}\right)  \, \left(\frac{\tau_{b}}{\tau_\Delta}\right) \, \left(\frac{ {\rm s}}{\tau_{edd}} \right).
\end{equation}
This value is maximally saturated near $10^{-6}$, suggesting a maximal peak luminosity near $10^{60} \, {\rm eV / s}$, although this value may also be smaller.

It is worth mentioning that Ref.~\cite{Rosa:2017ury} argued for the existence of a maximum black hole mass $M_c \sim 10^{-2} \, M_\odot$, above which the parametric resonance would not occur. This argument, however, was derived for the QCD axion, and does not generically apply to axion-like particles. Here, we focus on astrophysical (rather than primordial) black holes, and thus the remainder of this discussion applies only to axion-like particles.

Finally, a comment is in order regarding the environmental dependence of these events. Photons acquire an effective mass in the IGM due to their interactions with the ambient plasma, given by the plasma frequency:
\begin{equation}
	\omega_p = \sqrt{\frac{4\pi\alpha n_e}{m_e}} \, ,
\end{equation}
where $n_e$ is the electron number density. The plasma frequency in IGM at redshifts $z \lesssim 20$ has typical values of the order $\sim 10^{-14} \,$ eV, and thus axions of mass $m_a \lesssim 5 \times 10^{-13}$ eV are kinematically forbidden from decaying to photons, unless they reside in a large under density. For this reason, we choose to focus our attention on black holes with masses $M \lesssim 5 \times 10^{3} M_\odot$. It is important to bear in mind that black holes with masses near this threshold must reside in regions of the Universe where the baryon overdensity, $\Delta \equiv n_b / \bar{n}_b$,  is not enormously different from one. While black holes are of course expected to preferentially be found in over dense regions, simulations of black hole mergers have suggested that remanent black holes may be produced with velocities $\sim \mathcal{O}(5000) \,$ km/s~\cite{Campanelli:2007cga,Brugmann:2007zj,Healy:2008js,Lousto:2010xk,Lousto:2011kp,Gerosa:2014gja,Gerosa:2018qay,Sperhake:2019wwo} -- these `superkicks' are expected to occur when the spins of the merging black holes are antiparallel and lying in the orbital plane. Even larger kicks, potentially reaching $v \sim 0.1 $, may be generated should the merging black holes have hyperbolic encounters and the collision be ultra-relativistic~\cite{Sperhake:2010uv}. Black holes formed with such superkicks may easily escape their host galaxies, and if traveling at such speeds these objects will have highly suppressed rate of accretion, meaning it may be possible to escape their local environment before accumulating any additional matter. An alternative possibility is that the supernova responsible for the creation of the black hole itself may be capable of ejecting sufficient material to reduce the local plasma mass below the kinematic threshold.  Thus, the proposition of finding massive black holes in regions of $\Delta \sim 1$ does not seem unfathomable, although this problem presents an unavoidable uncertainty which may complicate an estimate of the net rate at which shockwaves are formed. 

Finally, the above concerns may be circumvented if the electromagnetic fields near the black hole are large. In this case, the photon dispersion relation can be modified, and the effect of the plasma mass can be dramatically reduced~\cite{kaw1970relativistic,max1971strong}. This was recently pointed out in the context of photon superradiance~\cite{Cardoso:2020nst}.

 \begin{figure}
	\includegraphics[width=0.48\textwidth,  bb=0 0 600 400]{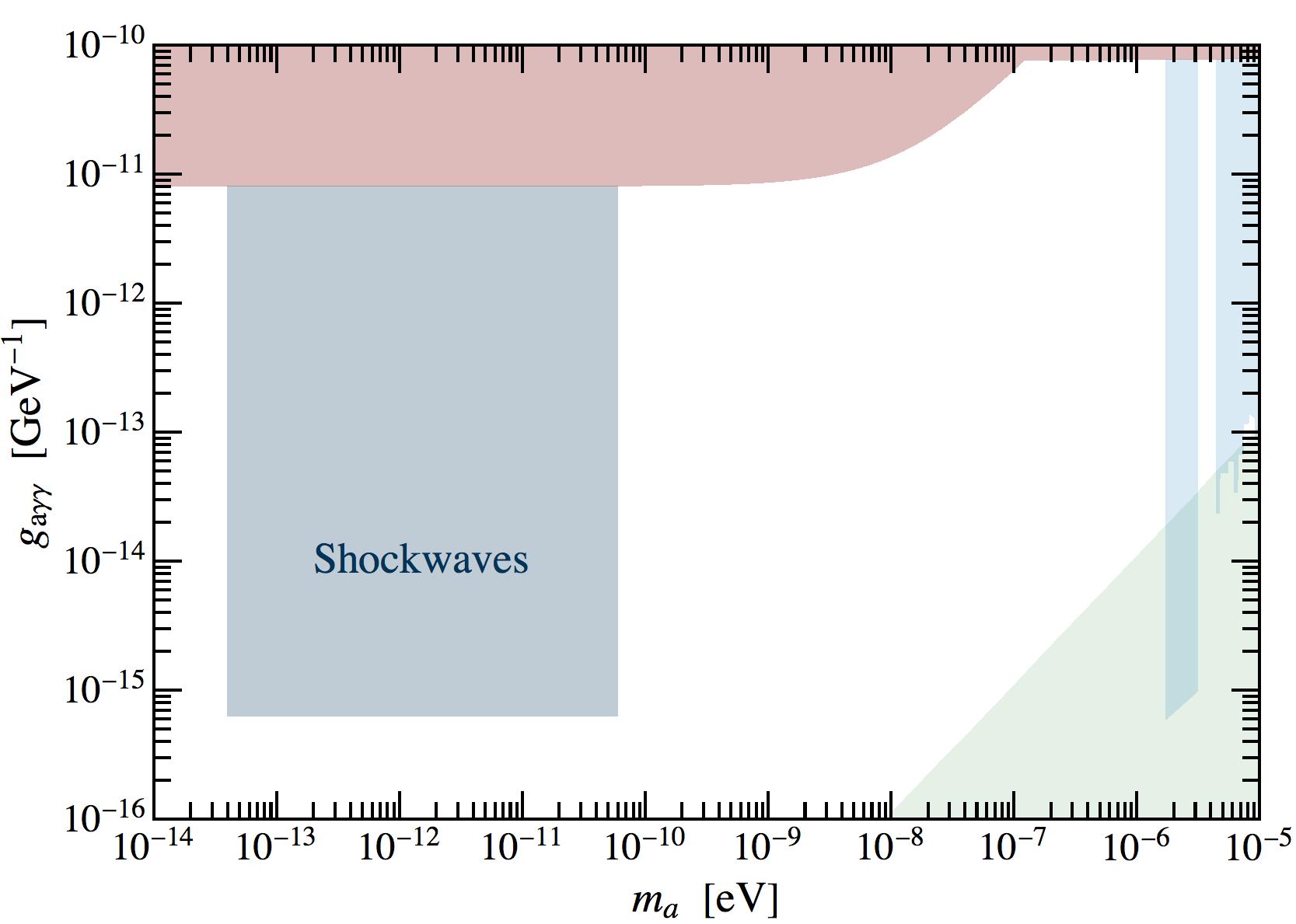}
    \caption{\label{fig:axion} Axion parameter space identified as capable of inducing shockwaves in the IGM. Shown for comparison are constraints from helioscopes~\cite{Anastassopoulos:2017ftl} and SN1987a~\cite{Payez:2014xsa} (red), and haloscopes~\cite{Hagmann:1998cb,Asztalos:2001tf,Du:2018uak} (light blue). The edge of the QCD axion band is highlighted in green in lower right corner~\cite{DiLuzio:2016sbl}.}
\end{figure}

\section{Shockwaves in the IGM} \label{sec:SW}

Axion clouds undergoing resonant decay can produce photons with short mean free paths in the IGM. Such low energy photons efficiently heat the gas via free-free absorption. The rate of absorption per unit volume is given by~\cite{draine2010physics}
\begin{equation}
	\Gamma_{ff} =  n_\gamma \, n_e \, \sigma_{ff} \, ,
\end{equation}
 with the free-free cross section in the limit $E_\gamma \ll T$
\begin{equation}
	\sigma_{ff} \simeq \frac{4 \pi^2 \alpha \sigma_T}{\sqrt{6\pi}} \, n_p \sqrt{\frac{m_e}{T}} \, \frac{g_{ff}(E_\gamma, T) }{T E_\gamma^2} \,  .
\end{equation}
Here, $T$ is the temperature of the gas, $n_p$ the proton number density, and $g_{ff}$ the gaunt factor, approximated here as~\cite{draine2010physics}
 \begin{equation}
 	g_{ff} \simeq 4.691 \left[1 - 0.118 \ln \left( \frac{\nu_\gamma}{10^{10} ( T / 10^4 \, {\rm K})^{3/2}} \right) \right] \, .
 \end{equation}
 For electron number densities and temperatures typical of the IGM, the mean free path of these particles, given by $\lambda_{mfp} = (n_e \times \sigma_{ff})^{-1}$, can be much smaller than parsec scales. If one assumes $\sim 10^{57}$ eVs of energy (\ie assuming $10^{60}$ eV/s imparted over a burst width of 1 ms\footnote{{ The luminosity is estimated here for a $10 \, M_\odot$ black hole with  $M_a / M \sim 10^{-7}$, and taking a duration given by twice the de Broglie wavelength of the axion cloud.}}) are instantaneously deposited in a parsec-sized volume, one expects the ambient particles to be heated to temperatures as high as $\sim 10^{12}$ K (in this estimation we have taken $z = 15$ and $x_e \sim 10^{-4}$, where $x_e \equiv n_e / n_H$ is the free electron fraction with $n_e$ and $n_H$ the electron and hydrogen number densities). The large pressure gradient induced from this heating  will induce an outward driven shockwave that sweeps through the IGM, heating and ionizing large regions of space; we attempt to quantify the effects of such a process here. 
 
 Before continuing, we identify the regions of parameter space for which a shockwave may develop. First, we are interested in considering astrophysical black holes, and thus we restrict attention to $M \gtrsim M_\odot$. For a given black hole mass, there will exist upper and lower limit on the axion mass, the former of which is set by requiring the superradiant condition is valid, and the latter by requiring the superradiance timescale is sufficiently small so as to extract energy. If the black hole mass is too large, axion decay may also be forbidden, and if the black hole is too small, the mean free path may be so large that energy deposition is not efficient enough to drive a shock wave (we set this limit to be the point where the total energy injection over the Eddington accretion timescale only amounts to $10^2$ eV / baryon, however observable effects likely require a much greater energy injection). We plot in \Fig{fig:mfp} the mean free paths of the photons produced as a function of black hole mass (for a fixed axion mass), assuming the event occurs either prior to reionization at $z = 15$ or post-reionization at $z = 6$. The regimes in which the superradiant decay will be impossible or inefficient are identified heuristically. Using \Fig{fig:mfp}, we identify the range of masses most likely to induce shock waves as: $4 \times 10^{-14} \, \lesssim m_a \lesssim 6 \times 10^{-11}$ eV. Using the condition on the axion-photon coupling given in \Eq{eq:ax_coup}, we can identify the parameter space in mass and axion-photon coupling potentially capable of generating large shockwaves -- this parameter space is highlighted in \Fig{fig:axion}, where we show for comparison current constraints from helioscopes~\cite{Anastassopoulos:2017ftl}, SN1987a~\cite{Payez:2014xsa}, and haloscopes (light blue)~\cite{Hagmann:1998cb,Asztalos:2001tf,Du:2018uak}, as well as the parameter space favored by the QCD axion~\cite{DiLuzio:2016sbl} (green).

In order to address the potential implications of a shockwave induced from the absorption of escaping radiation, we adopt a formalism that has been developed to trace the evolution of spherically symmetric explosions in the IGM, initially intended to treat supernovae winds from massive stars~\cite{Tegmark:1993td} (see \eg also~\cite{mckee1977theory,weaver1977interstellar,mccray1979violent,bruhweiler1980stellar,mccray1987supershells,Ostriker:1988zz} for early work on the subject). This treatment relies on the simple assumption that there exists a three phase medium, consisting of: (1) a dense shell of radius $R$ and thickness $\delta R$ containing a fraction $(1 - f_m)$ of the enclosed baryonic matter (assumed $f_m \ll 1$), (2) a uniform medium outside the shell with mean IGM density,  and (3) a hot isothermal plasma inside the shell. The shell will be driven outward by the pressure induced from the absorption and subsequent heating of the radiation escaping the black hole. As the shell expands, the plasma cools, ionizing the newly encountered neutral hydrogen and dissipating energy via Compton scattering with the CMB and bremsstrahlung emission. The medium may also undergo heating via collisions with the surrounding IGM. Eventually, the shell velocity will asymptote to match the rate of expansion of the Universe, and the internal temperature of the bubble will dissipate and asymptote to the background value.

Following~\cite{Tegmark:1993td}, the total mass in the shell at a time $t$ is given by $m(t) = \frac{4}{3}\pi R(t)^3 (1-f_m) \rho_b$, and the rate of change is simply
\begin{equation}
	\frac{\dot{m}}{m} =\frac{1}{R^3 \rho_b} \frac{d}{dt} \left( R^3 \rho_b \right) = 3 \left( \frac{\dot{R}}{R } - H\right) \, 
\end{equation}
if $\dot{R}/R > H$, otherwise 0 (implying the shell simply expands outwards with the Hubble flow). The shell will experience (1) a net braking force proportional to $(\dot{R} - H \dot{R}) \dot{m}$ associated with the acceleration of the material encountered, (2) an outward pressure force $4\pi R^2 p$ induced from the hot interior, and (3) a gravitational deceleration force $\frac{4}{3}\pi G R (\rho_{cdm} + \rho_b/2) + G M_{BH}/ R^2$, the latter term being small at the scales of interest. One can show the resultant equation for the radial motion of the blast wave is given by
\begin{equation}
	\dot{v} = \frac{8\pi G p}{\Omega_b^2 H^2 R \gamma^3} - \frac{3}{R}\left( v - H R\right)^2 - \left( \Omega_{cdm} + \frac{\Omega_b}{2}\right)\frac{H^2 R}{2} \, ,
\end{equation}
where the Lorentz factor $\gamma^3$ accounting for the relativistic suppression must be included to suppress the outward force, since the initial wave travels very near the speed of light.  The equation of state for the interior of the plasma is given by
\begin{equation}\label{eq:etop}
	E = \frac{3}{2} p \, \mathcal{V}_{sw} \, ,
\end{equation}
where $\mathcal{V}_{sw}$ is the volume contained in the shock bubble and $p$ the pressure. Energy conservation mandates
\begin{equation}
\frac{dE}{dt} = L - p \frac{d\mathcal{V}_{sw}}{dt} = L - p 4\pi R^2 v \, .
\end{equation}
The luminosity has contributions coming from: (1) the interior luminosity contributed by the superradiant event $L_{SR}$, (2) Compton cooling off CMB photons $L_{comp}$, (3) $L_{brem}$ thermal bremsstrahlung emission,  (4) $L_d$ heating from collisions between the shell and the IGM, and (5) $L_{ion}$ ionization of the surrounding medium. That is to say the net luminosity is given by
\begin{equation}
	L = L_{SR} + L_d - L_{comp} - L_{brem} - L_{ion} \, .
\end{equation}
We account for the superradiant luminosity using the time-averaged luminosity given in \Eq{eq:timeAL} and evolve the above equations from $t = 0$ to $t = \tau_{edd}$; after the superradiant energy injection has stopped, we convert the differential equations to redshift and use the final output of the time-based solutions as initial conditions for the cosmological evolution. We do not find significant changes to our results if the equivalent total energy is injected on over timescales $t \ll \tau_{edd}$. Notice that within these cosmological time scales, the heating of ions and of neutral hydrogen can indeed be treated as instantaneous, see e.g.  \cite{BlasWitte_photonSR}.

\begin{figure*}
	\centering
	\includegraphics[width=0.45\textwidth, bb=0 0 625 450]{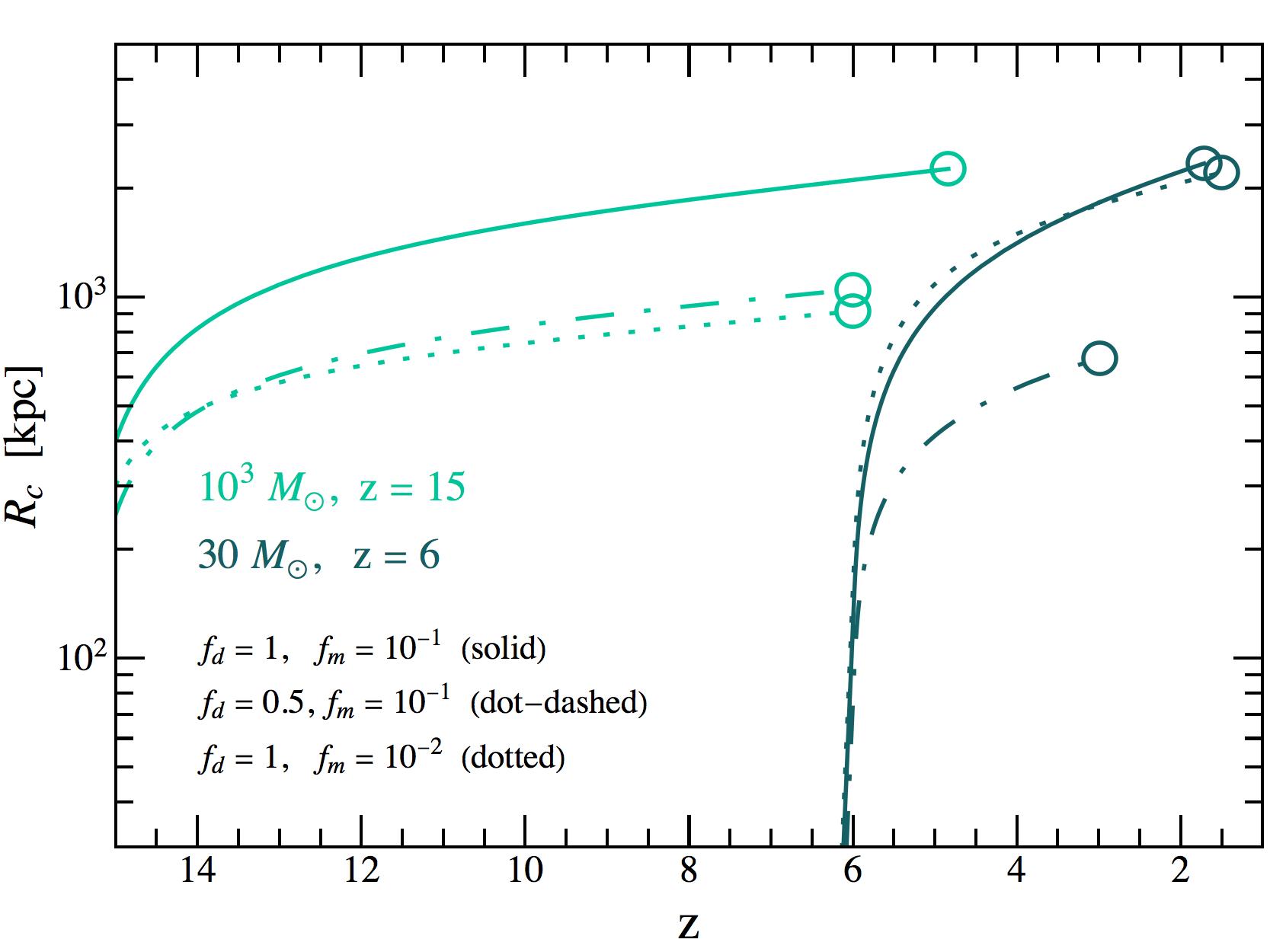}
	\includegraphics[width=0.45\textwidth, bb=0 0 625 450]{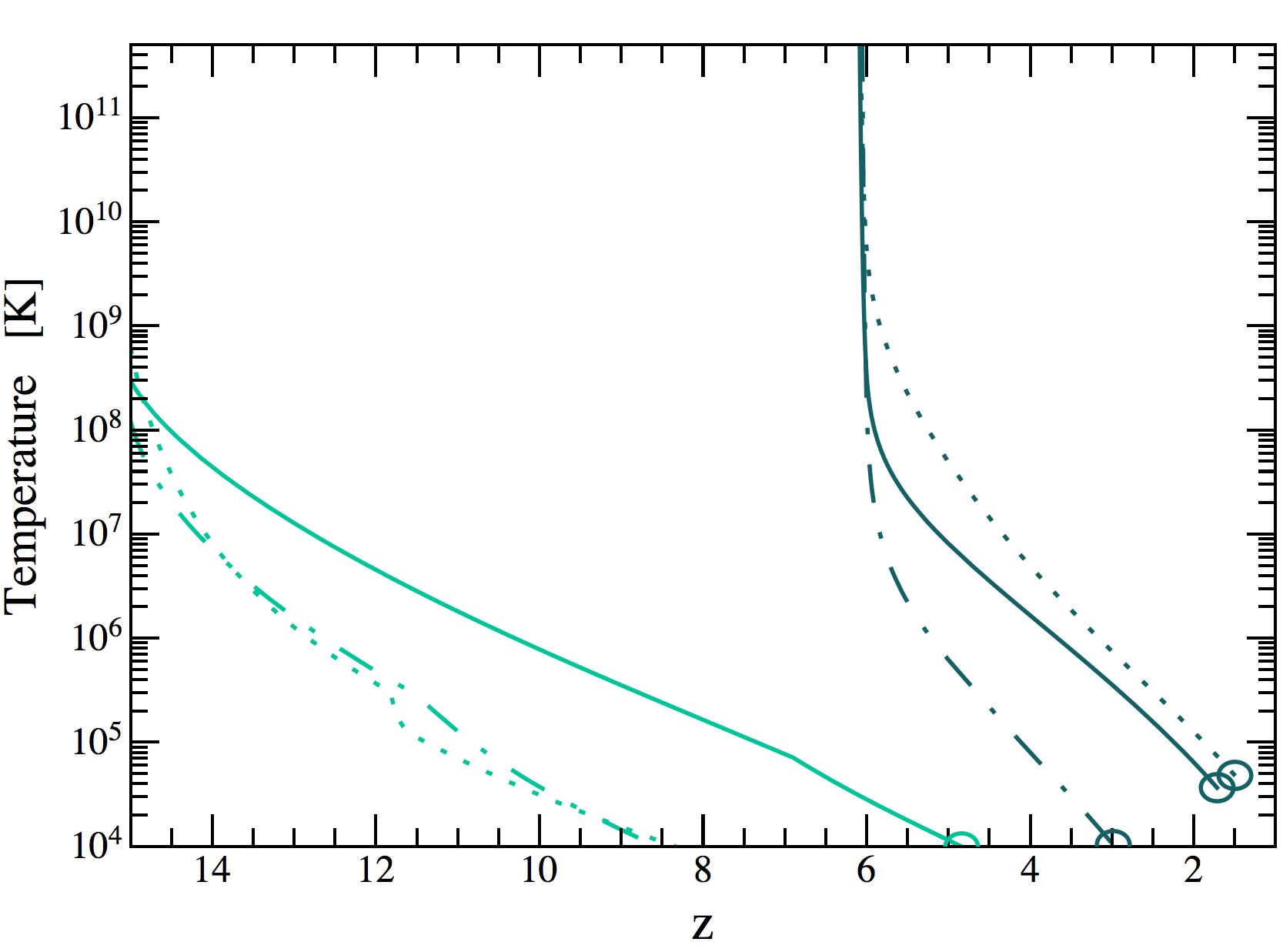}
	\includegraphics[width=0.45\textwidth, bb=0 0 625 450]{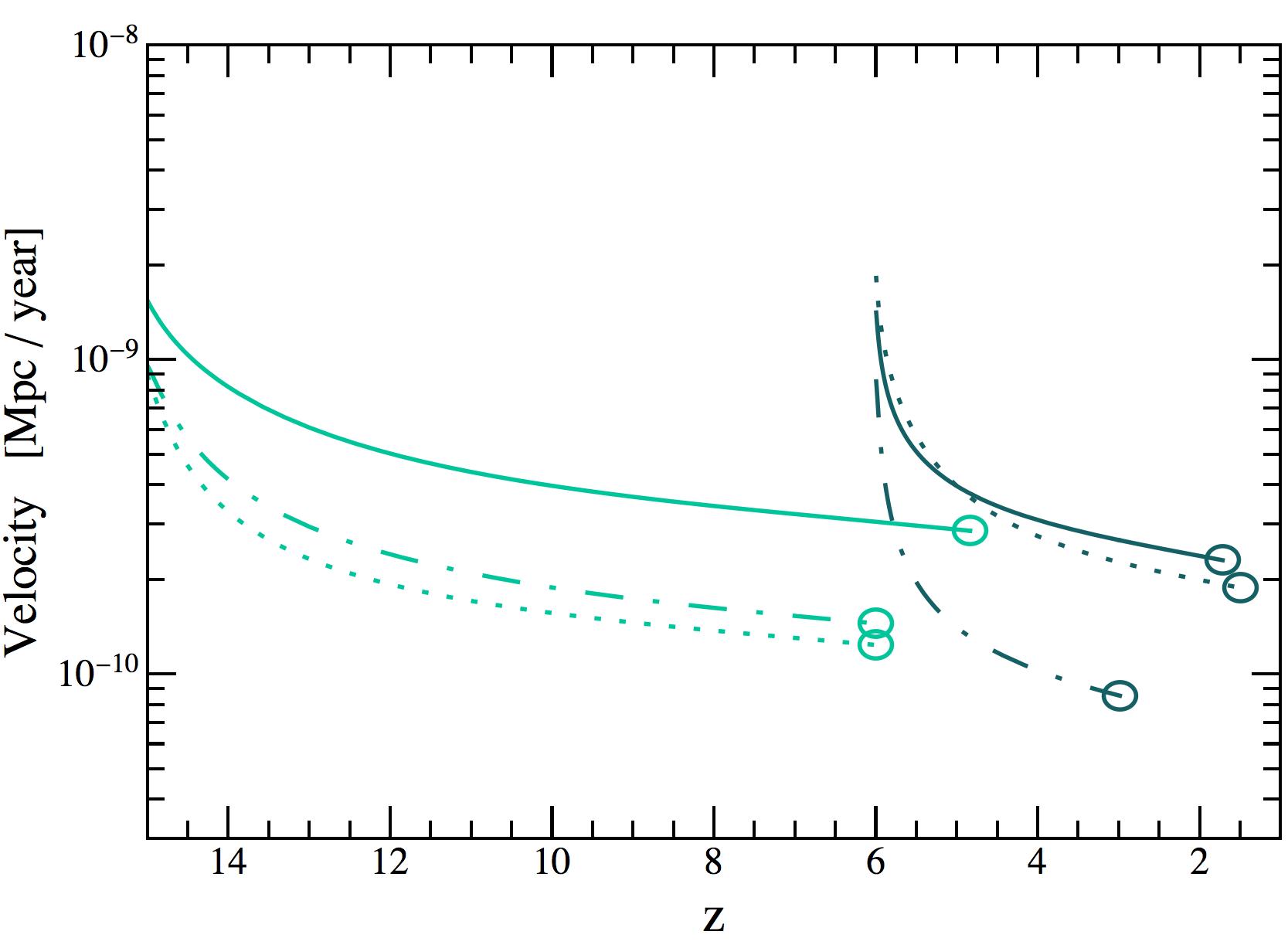}
	\includegraphics[width=0.45\textwidth, bb=0 0 625 450]{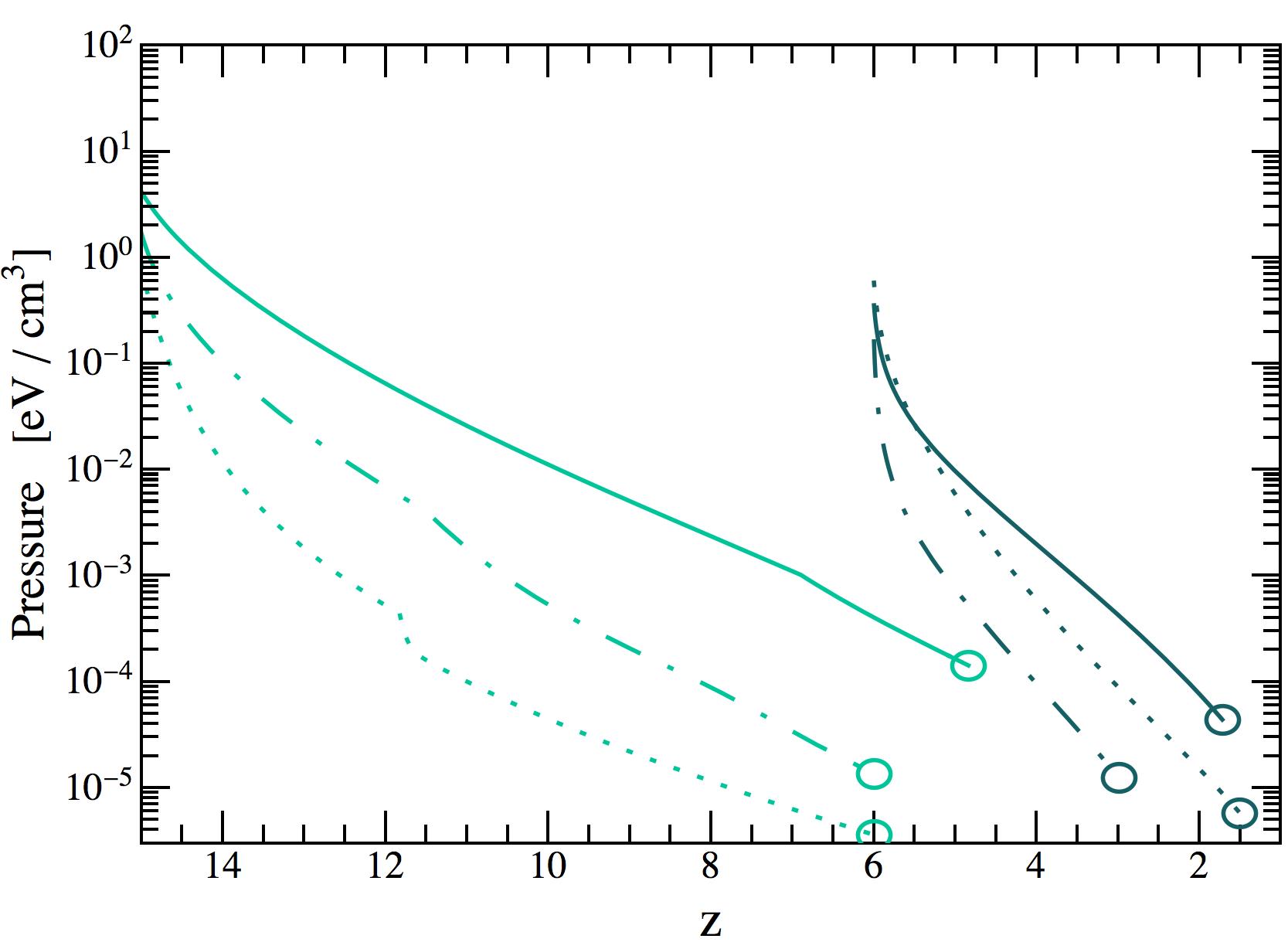}
	\caption{\label{fig:ShockWaveEx} Shockwave solutions generated for $M = 10^3 M_\odot$ and $30 M_\odot$, initiated at $z = 15$ and $6$, respectively. Results are shown for fiducial model of $f_d = 1$ and $f_m = 0.1$, and the effect of varying these parameters to $f_d = 0.5$ and $f_m = 10^{-2}$. All curves assume ${G} \, m_a M = 0.2$. }
\end{figure*}

The luminosity of Compton cooling of relativistic electrons off CMB photons is given by
\begin{equation}
	L_{comp} = \frac{4}{3} \beta^2 \gamma^2 \sigma_T \, \frac{\pi^2}{15} \, T_\gamma^4 \, n_e \, \mathcal{V}_{sw}
\end{equation}
where $T_\gamma$ is the temperature of the CMB, and that of bremsstrahlung cooling is
\begin{equation}
	L_{brem} = 1.422\times10^{-25} \sqrt{T_g / 10^4 \, {\rm K}} \, n_p \, n_e \, g(T) \, \mathcal{V}_{sw} \, ,
\end{equation}
where $T_g$ is the temperature of the gas and $g(T)$ is the gaunt factor.  Next, we need to address the fate of the energy created in the inelastic collisions between the shell and the IGM. This energy could be either radiated away in a shock cooling process or heat the shell (and subsequently the interior plasma). We adopt a conservative approach to deal with this by parameterizing the uncertainty in how much energy gets radiated / reabsorbed via a constant $f_d \in [0, 1]$, with the limit that $f_d \rightarrow 1$ representing the limit of total energy retention. The net luminosity for this process is then given by
\begin{equation}
	L_{d} = f_d \frac{3 m}{2 R} \left( v - H R\right)^3 \, .
\end{equation}
Lastly, the ionization luminosity simply accounts for the energy required to ionize the neutral hydrogen absorbed by the expanding wave, and is given by 
\begin{equation}
	L_{ion} = f_m\, f_{x_e}\, n_b \, I_H \, 4\pi \, R^2 \left( v - H \, R \right) \, ,
\end{equation}
where $I_H = 13.6$ eV is the ionization threshold of hydrogen, and $f_{x_e}$ is the fraction of neutral hydrogen. We assume $f_{x_e}$ is the conventional {\emph{tanh}} function used to describe reionization, with a central reionization redshift taken to be $z_{re} =7$ and width $\Delta z = 0.5$. Finally, the evolution of the internal pressure can be related to the net luminosity and the expansion of the shell via
\begin{equation}
	\frac{dp}{dt} = \frac{L}{2\pi R^3} - 5 \frac{v}{R} p \, .
\end{equation}

With this formalism in hand, we restrict our attention to two scenarios: a $10^3 M_\odot$ black hole at $z = 15$, and a $30\, M_\odot$ black hole at $z = 6$, both with an axion mass $m_a \sim 0.2 / ({G}\,M)$. In our fiducial model, we take $f_d = 1$ and $f_m = 0.1$, however we also consider $f_d = 0.5$ and $f_m = 10^{-2}$ to illustrate the effect of changing these parameters. The results, showing the comoving radius $R_c = (1+z) R$, velocity, pressure, and temperature evolution of each shockwave are shown in \Fig{fig:ShockWaveEx}. We evolve the system until the temperature of the bubble falls below that of the IGM. In the pre-reionization scenario, we assume the temperature of the external medium is given by that of $\Lambda$CDM, while lower redshifts we model the IGM temperature as~\cite{Sunyaev:2013aoa}
\begin{equation}
	T = \begin{cases}
		10^4 \, \hspace{.1cm} {\rm K}  & z > 3 \\
		10^6 / (1+z)^{3.3} \, \hspace{.1cm} {\rm K}   & z \leq 3 \, .
	\end{cases}
\end{equation}
Under reasonable assumptions, these expanding pressure bubbles seem to be capable of expanding to comoving distances as large as $\sim \mathcal{O}(1)$ Mpc, and thus may leave observable signatures in the form of heating and ionization.

\section{Observable Features} \label{sec:obs}

Finally,  let us turn our attention to the observational consequences of these shockwaves. The dominant impact is a heating, and ionization in the case of the $z = 15$ superradiance, of the IGM over distance scales potentially as large as $\mathcal{O}(1)$ Mpc. While this may leave strong signatures \eg in the Ly-$\alpha$ forest and the evolution of structure, these signatures often rely on averaging over some set of statistics. Given the large uncertainties in the rate and distribution of black holes undergoing superradiance, we would prefer to focus our attention on probes that may identify individual superradiant events, with the understanding that prospects may improve should the rate be larger. We believe the strongest of such signatures arises from angular features in the optical depth and spectral distortions in the CMB spectrum. 

After recombination, heated gas can induce $y$-type spectral distortions in the CMB, arising from the effect of Compton scattering. This effect is often parameterized via the so-called $y$-parameter, given by~\cite{Sunyaev:2013aoa}
\begin{equation}
	y_c = \int \, dz \, \frac{(T - T_{\rm cmb})}{m_e } \, \frac{ \sigma_T \, n_e}{H(z) \, (1+z)} \, .
\end{equation}
It is well-known that the hot IGM after reionization is expected to imprint a spectral distortion at the level of $y_c \sim 3 \times 10^{-7}$~\cite{Zeldovich:1969ff,Hu:1994bz}. The distortions introduced by hot gas in clusters may be larger, closer to $2\times 10^{-6}$, which may further complicate the identification of such effects (see \eg~\cite{Chluba:2019nxa}). Thus, if the generated shockwave can heat the IGM to a sufficiently high level (and over a sufficiently large angle in the sky), it may be possible to produce strong signals on top of the astrophysical background.

\begin{figure*}
	\includegraphics[width=0.48\textwidth, bb=0 0 625 450]{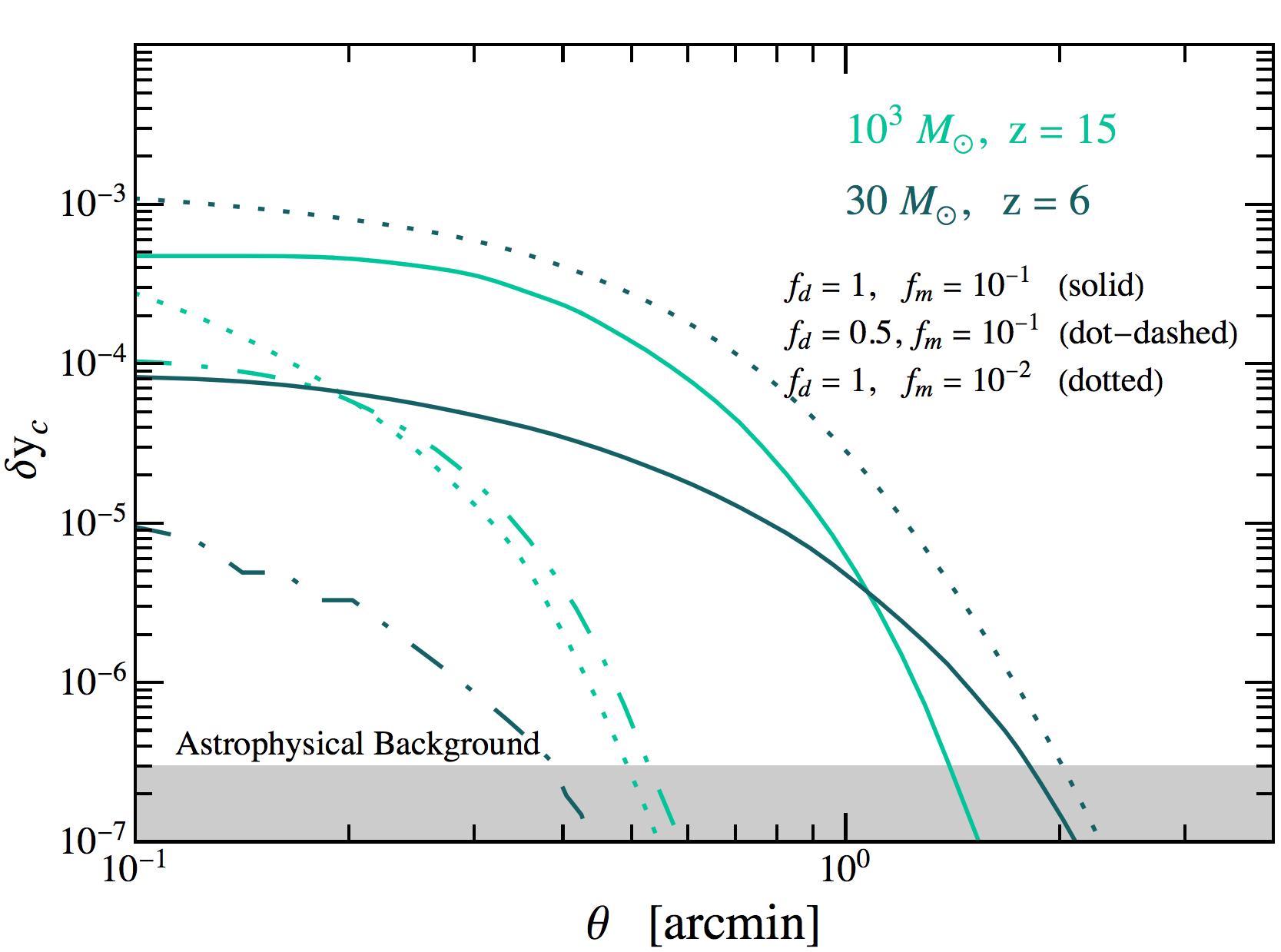}
	\includegraphics[width=0.48\textwidth, bb=0 0 625 450]{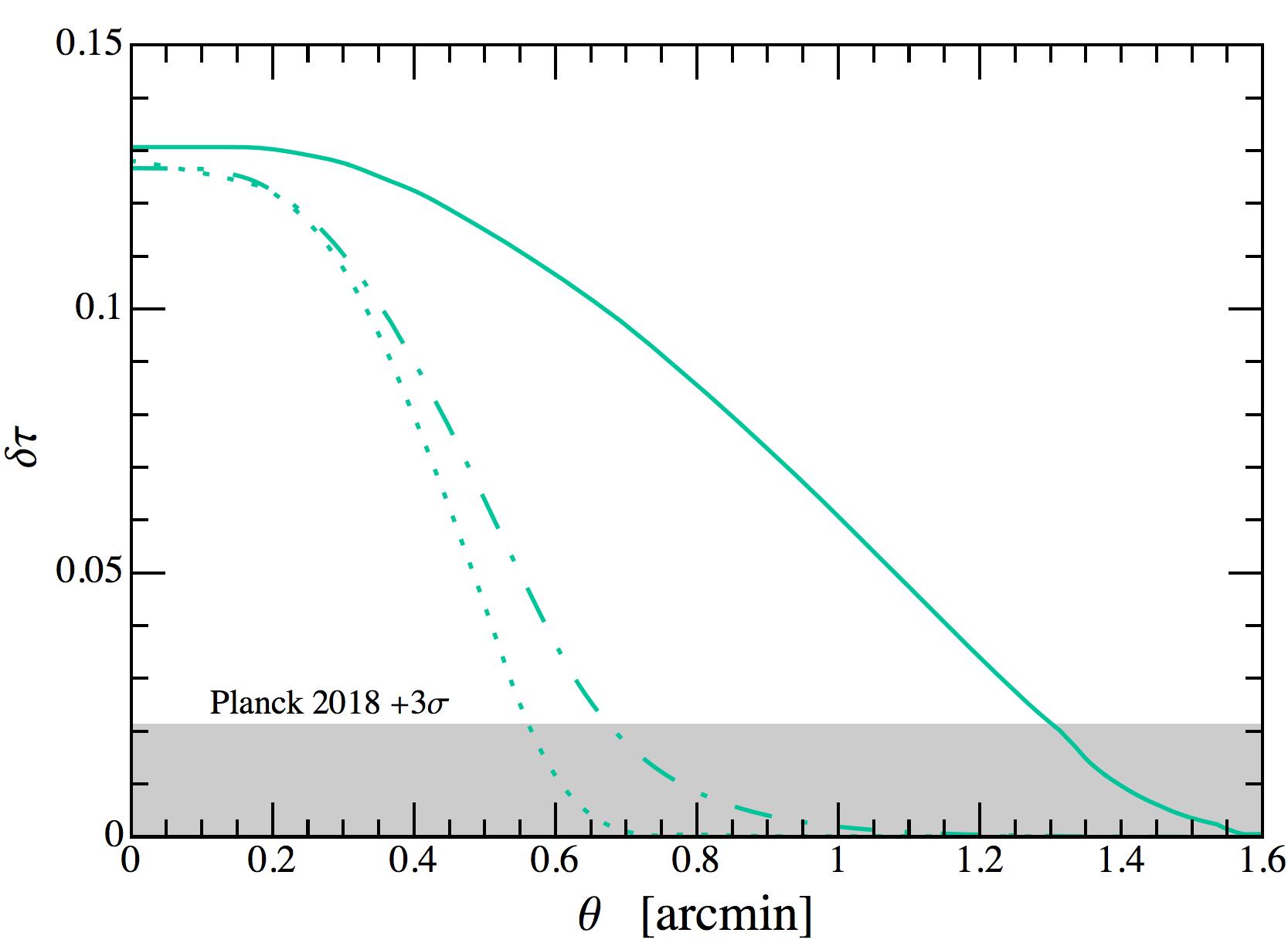}
	\caption{\label{fig:delY_c} Relative contribution to Compton $y-$parameter (left), defined as $\delta y_c \equiv y_c - y_{c, \, \Lambda{\rm CDM}}$, and optical depth (right) as a function of angular scale. The $3\sigma $ upper limit on the optical depth from Planck, based on the sky-average value, is shown as a reference. For comparison, the angular resolution of PRISM~\cite{Andre:2013afa,Andre:2013nfa} spans between $10^{-1}$ and $17$ arc-minutes, depending on the frequency. }
\end{figure*}

We analyze the potential signal by looking at the relative difference between the spectral distortion induced from the shockwave and that expected from astrophysical sources. The $\Lambda$CDM contribution is modeled assuming the second ionization of helium occurred at $z = 3.5$ with a width $\delta z = 0.5$, and the temperature of the IGM is described in the previous section. This resultant difference in spectral distortions is given by $\delta y_c =  y_{c}^{\rm SR} - y_c^{\Lambda{\rm CDM}}$, and is shown in the left panel \Fig{fig:delY_c} as a function of angular degree on the sky. The leading constraints to date come from FIRAS, having measured the $y-$parameter to the level of $y_c \lesssim 10^{-5}$; FIRAS~\cite{Fixsen:1996nj}, however, had an angular resolution of $\sim 10^\circ$, and thus would not have been capable of observing the induced heating from axion superradiance (unless the rate of events is quite large). Proposed spectral distortion experiments such as PIXIE~\cite{Kogut:2011xw}, PRISM~\cite{Andre:2013afa,Andre:2013nfa}, or the proposals for Voyage 2050~\cite{Basu:2019rzm,Delabrouille:2019thj,Chluba:2019nxa} have angular resolutions as low as $\mathcal{O}(5)$ arc-seconds, and have estimated sensitivities near the level of $y_c \sim 10^{-7}$; consequently \Fig{fig:delY_c} shows that photon superradiance has the potential to induce strong signals observable in these experiments. 

The ionization induced by superradiant events that are triggered prior to reionization can also leave an imprint in the optical depth. Inhomogeneous features in the optical depth alter the statistical features in the CMB via the $(i)$ screening of anisotropies (i.e. temperature and polarization anisotropies are multiplied by a factor of $e^{-\tau(\hat{n})}$), $(ii)$ generation of polarization via Compton scattering, and $(iii)$ the kinetic Sunyaev-Zel'dovich (kSZ) effect. The shockwaves identified here may naturally appear as strong anomalies in future efforts to detect the signatures from patchy reionization (see \eg~\cite{Gruzinov:1998un,Zahn:2005fn,Dvorkin:2008tf,Namikawa:2017uke,Roy:2018gcv} for detection of patchy reionization in the CMB). The line of sight contribution to the optical depth from a single superradiant event  is given by
\begin{equation}\label{eq:del_tau}
	\delta \tau = \int \, dz \frac{1}{H(z) (1+z)} \left[ n_e(z) - n_{e,0}(z)\right] \sigma_T \, ,
\end{equation}
where $n_e$ is the free electron fraction in the superradiant model and $n_{e,0}$ is that from the default cosmological model; we take the default reionization history to be given by the conventional {\emph{tanh}} reionization history, centered at $z=7$ with width $\Delta z = 0.5$.  The $3\sigma$ upper limit from \cite{Aghanim:2018eyx} is shown for comparison. Telescopes measuring small scale features in the optical depth may be capable of identifying such features.

Thus far our discussion has been focused on the effects arising from axion superradiance around a single black hole. Needless to say, realistic population models predict many  black holes capable of efficiently inducing axion superradiance, and thus one may expect large scale features to appear; if the number of events is sufficiently large, it may be possible that FIRAS and Planck have even already begun to probe interesting regions of parameter space. We can estimate the approximate number of  events required to produce all-sky features simply by looking at the ratio of the fractional sky coverage of a single event; for $\sim 1$ arc-minute scales, this fraction amounts to a factor of $\sim 10^{-9}$. Cosmologically, the existence of a population of $10^9$ rapidly rotating black holes in the mass range identified in \Fig{fig:mfp} is not unreasonable, and thus a dedicated study of the resonant decay of superradiant axion clouds using black hole populations would be  of great interest. We leave this to future work.

\section{Conclusions} \label{sec:conc}
 
Light axions can generate enormous clouds around rapidly spinning black holes via a phenomenon known as superradiance. For sufficiently large axion-photon couplings ($g_{a\gamma\gamma} \gtrsim 10^{-15} \, {\rm GeV^{-1}}$), these clouds may undergo resonant decay into low energy monochromatic photons that propagate into the IGM. In this work we have investigated the observable signatures that may arise in the CMB from the superradiant production and subsequent resonant decay of axions. 

The photons produced in such processes must have very low energies, no more than a few orders of magnitude above the plasma frequency of the surrounding medium. Consequently, these photons are efficiently absorbed via inverse bremsstrahlung scattering off ambient electrons and ions, with typical mean free paths capable of being much below the parsec scale.  The medium is immediately heated to ultra-high temperatures, producing a huge pressure gradient that can drive shockwaves to Mpc scales. This process heats and ionizes the IGM, and can leave an imprint on both the optical depth and y-type spectral distortions. For a single event, this is expected to produce an anisotropic feature on arc-minute scales, below the observable limit but potentially detectable with future experiments. For couplings $g_{a\gamma\gamma} \gtrsim 10^{-13} \, {\rm GeV^{-1}}$, spectral distortions may also arise from  the conversion of CMB photons to axions, potentially providing a complementary probe~\cite{Mukherjee:2018oeb,Mukherjee:2018zzg,Mukherjee:2019dsu}.

Accounting for realistic populations of black holes may dramatically enhance the observability of this signature, however the uncertainty in the distribution of black holes at high redshifts is large -- we leave a more thorough investigation of the effects of black hole populations to future work. Finally, we emphasize that if a putative detection of either feature be observed, it may be possible to gain direct insight into the axion mass, the distributions of highly spinning black holes, and the properties of the IGM.  Resonant decay of axion superradiance thus offers an interesting and novel probe with importance to particle physics, astrophysics, and cosmology alike.

~\\ \noindent{\it \bf Acknowledgments:} 
The authors thank Vivian Poulin, Jens Chluba, Thomas Schwetz, Vitor Cardoso, and Paolo Pani for their comments. SJW acknowledges support under the Juan de la Cierva Formaci\'{o}n Fellowship.

\bibliography{biblio}

\end{document}